\documentclass[twocolumn,aps,pra,showpacs,groupedaddress,superscriptaddress
]{revtex4-1}

\usepackage[utf8]{inputenc}  
\usepackage[T1]{fontenc}

\usepackage[normalem]{ulem}

\usepackage[dvipsnames]{xcolor}
\usepackage{graphicx,epsfig}
\usepackage{braket}
\usepackage{color}
\usepackage{amsmath}
\usepackage{amsfonts}
\usepackage{fancyhdr}
\usepackage{mathrsfs}
\usepackage{mathtools}

\usepackage[pdftex,linkcolor=blue,citecolor=blue,urlcolor=blue,colorlinks]{hyperref}

\newcommand{\mean}[1]{\langle#1\rangle}

\newcommand{\dgr}{^{\dagger}}

\DeclarePairedDelimiter\abs{\lvert}{\rvert}%
\DeclarePairedDelimiter\norm{\lVert}{\rVert}%
\makeatletter
\let\oldabs\abs
\def\abs{\@ifstar{\oldabs}{\oldabs*}}
\let\oldnorm\norm
\def\norm{\@ifstar{\oldnorm}{\oldnorm*}}
\makeatother

\renewcommand{\vec}[1]{\mbox{\boldmath$#1$}}

\newcommand{\enum}[1]{\mathit{e}^{#1}}

\newcommand{\ham}{\hat{H}}
\newcommand{\Spvek}[2][r]{%
  \gdef\@VORNE{1}
  \left(\hskip-\arraycolsep%
    \begin{array}{#1}\vekSp@lten{#2}\end{array}%
  \hskip-\arraycolsep\right)}
\newcommand{\appropto}{\mathrel{\vcenter{
\offinterlineskip\halign{\hfil$##$\cr
  \propto\cr\noalign{\kern2pt}\sim\cr\noalign{\kern-2pt}}}}}

\newcommand{\overbar}[1]{\mkern 1.5mu\overline{\mkern-1.5mu#1\mkern-1.5mu}\mkern 1.5mu}

\DeclareMathOperator{\sgn}{sgn}

\newcommand{\fref}[1]{Fig.~\ref{#1}}

\newcommand{\sref}[1]{Sec.~\ref{#1}}

\begin{abstract}
Excited-state quantum phase transitions depend on and reveal the structure of the whole spectrum of many-body systems. While they are theoretically well understood, finding suitable signatures and detect them in actual experiments remains challenging. For instance, in spinor gases, excited-state phases have been identified and characterized through a topological order parameter that is challenging to measure in experiments. Here, we propose the Raman-dressed spin-orbit coupled gas as a novel platform to explore excited-state quantum phase transitions. In a weakly-coupled regime, the dressed system is equivalent to a spinor gas with tunable spin-spin interactions. Through this equivalence we are able to define a new excited-state phase of the dressed gas. The phase is characterize by the the behavior of the spatial density modulations, or stripes, induced by spin-orbit coupling, and can in principle be measured in current state-of-the-art experiments with ultracold atoms. Conversely, we show that the properties of the excited phase can be exploited to prepare stripe states with large and stable density modulations.
\end{abstract}

\begin{document}
\title{Excited-state quantum phase transitions in spin-orbit coupled Bose gases}
\author{J. Cabedo}
\affiliation{Departament de F\'isica, Universitat Aut\`onoma de Barcelona, E-08193 Bellaterra, Spain.}
\author{A. Celi}
\affiliation{Departament de F\'isica, Universitat Aut\`onoma de Barcelona, E-08193 Bellaterra, Spain.}
%\email{Josep.Cabedo@uab.cat} %optional
%\date{\today}


\maketitle

\section{Introduction}\label{Sec-intro}

Harnessing quantum matter with light is at the heart of quantum technology \cite{Gardiner-2015,Celi-2017}. Artificial spin-orbit coupling (SOC) in ultracold atom gases is a prominent example \cite{Dalibard-2011, Goldman-2014, Maciej-Book-2012}. Spinor gasses dressed by Raman coupling \cite{Lin-2009,Lin-2011} interact differently \cite{Williams-2012},  host stripe phases \cite{Li-2017,Putra-2020} with supersolid-like properties (see also \cite{Hou-2018}, for dipolar gases realization see \cite{Tanzi-ss-prl-2019, Bottcher-prx-2019, Chomaz-prx-2019}), or even realize a topological gauge theory \cite{Ramos-2021}. Here we propose to use Raman-dressed spinor-orbit-coupled gasses for studying dynamical \cite{Heyl-rpp-2018} and excited \cite{Cejnar-jphysA-2021} quantum phase transitions in spinor Bose-Einstein condensates (BECs).

%Harnessing quantum matter with light is at the heart of quantum technology \cite{Gardiner-2015,Celi-2017}. Artificial spin-orbit coupling (SOC) in ultracold atom gases is a prominent example \cite{Dalibard-2011, Goldman-2014, Maciej-Book-2012}. Since its landmark realization in Raman-dressed condensates \cite{Lin-2009,Lin-2011}, ultracold spin-orbit coupled gases have given rise to a large body of research \cite{Wang-prl-2010, Ho-prl-2011,  Li-2013, Zhang-2016}. Spinor gasses dressed by Raman coupling  exhibit many interesting properties. For instance, the interplay between synthetic SOC and inter-atomic interactions can result into anomalous scattering processes \cite{Williams-2012}, or even a topological gauge theory \cite{Ramos-2021}. In ultracold Bose gases, the exotic combination of SOC and superfluidity has drawn especial attention. These systems have been shown to host stripe phases with supersolid-like properties \cite{Li-2017,Putra-2020} (see also \cite{Hou-2018}, for dipolar gases realization see \cite{Tanzi-ss-prl-2019, Bottcher-prx-2019, Chomaz-prx-2019}). Here we propose to use Raman-dressed spinor-orbit-coupled gasses for studying dynamical \cite{Heyl-rpp-2018} and excited \cite{Cejnar-jphysA-2021} quantum phase transitions in spinor Bose-Einstein condensates (BECs).

In analogy to ground-state quantum phase transitions \cite{Vojta-rpp-2003,Sachdev_QPT_2nded-2011}, dynamical and excited-state quantum phase transitions involve the existence of singularities, respectively, in the time evolution and in the energy (or an order parameter) of an excited energy level, and can extend across the excitation spectra. Dynamical phase transitions have been demonstrated in quench experiments with cold atoms in optical lattices \cite{Flaschner-natphys-2017, Sun-prl-2018, Smale-science-2019} and cavities \cite{Muniz-nature-2020}, trapped ions \cite{Jurcevic-prl-2017, Zhang-nature-2017}, and with superconducting qubits \cite{Xu-sciadv-2020}. At the same time, excited-state quantum phase (ESQP) transitions  have been shown to occur in a variety of models \cite{Leyvraz-prl-2005, Ribeiro-prl-2007, Caprio-ESQPTs-2008, Brandes-pre-2013, Stransky-aop-2014, Opatrny-sr-2018, Macek-prc-2019}, and have been observed in superconducting microwave Dirac billiards \cite{Dietz-prb-2013}. Recently, dynamical and ESQP transitions have been theoretically \cite{Ceren-pra-2018, Feldmann-prl-2021} and experimentally \cite{Yang-PRA-2019, Tian-PRL-2020} studied in spin-1 BECs with spin-changing collisions. 

In \cite{Cabedo-stripe-arxiv-2021}, we showed that the Raman-dressed spin-1 SOC gas at low energy can be understood as an \emph{artificial} spin-1 gas with \emph{tunable} spin-changing collisions that can be adjusted with the intensity of the Raman beams. For weak Raman couplings and zero total magnetization, the dressed system is well described by a one-axis-twisting collective spin Hamiltonian \cite{Kitagawa-1993,Law-1998,Duan-2000}. The realization of the same model in undressed spinor condensates has led to the observation of various quantum many-body phenomena \cite{Stamper-Kurn-2013}, including the formation of spin domains and topological defects \cite{Stenger-nature-1998, Sadler-2006, Bookjans-prl-2011, Vinit-2013, Hoang-natcoms-2016, Anquez-prl-2016, Prufer-nature-2018, Chen-prl-2019, Kang-prl-2019, Jimenez-Garcia-2019, Prufer-natphys-2020}, and the generation of macroscopic entanglement \cite{, Bookjans-2011, Lucke-science-2011, Gross-nature-2011, Hamley-nature-2012, Zhang-prl-2013, Gabbrielli-prl-2015, Peise-ncomms-2015, Hoang-PNAS-2016,  Luo-2017, Zou-2018, Kunkel-science-2018, Pezze-prl-2019,  Qu-prl-2020}, with prospects for metrological applications \cite{Pezze-review-2018}. 

The map to pseudospin degrees of freedom (see \fref{Fig_eff_low_en_phys}) highlights the potential advantage of SOC dressed gases for engineering quantum many-body physics: the enhanced tunability of the system and the built-in entanglement between the emerging collective spin structures and the orbital degrees of freedom. In this work we employ these unique features to identify a novel Excited-Stripe (ES) phase of the spin-1 SOC gas. The phase is in correspondence to the Broken-Axisymmetry (BA') excited phase of the effective collective spin model, discussed in \cite{Feldmann-prl-2021}, which is characterized by a topological order parameter, and can extend over the whole spectrum of the Hamiltonian. In the SOC gas, ES phase comprises the classical phase-space trajectories with nonzero time average of the spatial modulations of the density of the gas. 

We exploit the relationship between the topological order parameter and the stability of the density modulations in the SOC gas to design a novel detection protocol for the ESQPs of the spinor gas. In the dressed gas, having an interferometer built-in generated by SOC makes a measurement of the contrast of the stripe equivalent to a simultaneous measurement of the amplitude and phase of the dressed spin components. Remarkably, this approach benefits from an intrinsic robustness to magnetic fluctuations, which constraint the current proposals for detectiong the excited phases of the model in spinor gases with {\it intrinsic} spin changing collisions \cite{Feldmann-prl-2021}. 

Finally, through the effective model, we are able to provide a robust protocol to prepare striped states. The ES phase of the gas can be accessed from an initially unpolarized gas via crossing an ESQP transition in a {\it two-step quench scheme}. With such approach, we show that the ES phase can be realized in current state-of-the-art experiments with spin-1 SOC gases, with the prepared states exhibiting large and stable density modulations. At the same time, the proposal introduces a novel procedure to access the striped regime of the spin-1 with SOC, which as ground-state phase has a very narrow region of stability \cite{Martone-2016} and it has yet to be experimentally demonstrated.

The paper is organized as follows. In \sref{Sec-effective-model} we briefly review the Raman-dressed spin-1 gas and its description as a collective pseudo-spin Hamiltonian with tunable spin interactions. In \sref{Sec-ESphase}, we introduce the novel ES phase of the dressed condensate, and show that its experimental signature can provide a new means to detect the ESQP transitions of the collective spin model. In \sref{Sec-quenchES} we propose a robust protocol to prepare ES states, which we benchmark in \sref{Sec-experiment}. Finally, we briefly recap and draw our conclusions in \sref{Sec-conclusions}.

\section{Raman-dressed gas as an artificial spinor gas}\label{Sec-effective-model}

We consider a spin-1 BEC of $N$ atoms of mas $m$ subject to synthetic SOC with equal  Rashba and Dresselhaus contributions, as experimentally realized via Raman-coupling two Zeeman pairs $\{\ket{1,1},\ket{1,0}\}$ and $\{\ket{1,0},\ket{1,-1}\}$ independently, as in \cite{Campbell-2016}. In the presence of dressing, the kinectic Hamiltonian can be written as
\begin{equation}\label{eq_kinetic_ham}
\hat{\mathcal{H}}_{\mathrm{k}} = \frac{\hbar^2}{2m}\left(\vec{k} - 2k_r \hat{F}_z \vec{e}_z\right)^2 + \frac{\Omega}{\sqrt{2}}\hat{F}_x + \delta \hat{F}_z + \epsilon \hat{F}_z^2,    
\end{equation} 
where $\hbar\hat{F}_j$ are the spin-1 matrices. Here, $\Omega$ quantifies the Raman coupling strength. By simultaneously adjusting the detuning from resonance of each Raman pair, the strengths of an effective quadrupole tensor field and a magnetic field term, $\epsilon$ and $\delta$, respectively, can be independently tuned in the laboratory (see methods from \cite{Campbell-2016}).

The many-body scenario for the weakly-interacting gas in mean-field regime is captured by the  energy functional
\begin{equation}\label{eq_energy_functional}
E[\vec{\psi}] = \vec{\psi}^{*}\!\! \left(\hat{\mathcal{H}}_{\mathrm{k}} \!+\! V_\mathrm{t}\right) \vec{\psi} + \frac{g_s}{2}\abs{\vec{\psi}}^4 \!+\! \frac{g_a}{2}\sum_j (\vec{\psi}^{*} \hat{F}_j \vec{\psi})^2 ,
\end{equation}
where  $\hat{\vec{\psi}} = (\hat{\psi}_{-1},\hat{\psi}_{0},\hat{\psi}_{1})^T$ is the spinor condensate wavefunction, normalized to the total number of particles as $\int d\vec{r} \vec{\psi}\dgr \vec{\psi} = N$. The spin-symmetric and non-symmetric interaction couplings are given by $g_s = 4\pi \hbar^2(a_0+ 2a_2)/3m$ and $g_a = 4\pi \hbar^2(a_2-a_0)/3m$, where $a_0$ and $a_2$ are the scattering lengths in the $F=0$ and $F=2$ channels, respectively. For simplicity, we will consider that the gas is confined with an isotropic harmonic potential $V_\mathrm{t} = \frac{1}{2}m\omega_\mathrm{t}^2 \vec{r}^2$. 

\begin{figure}[t]
\includegraphics[width=0.99\linewidth]{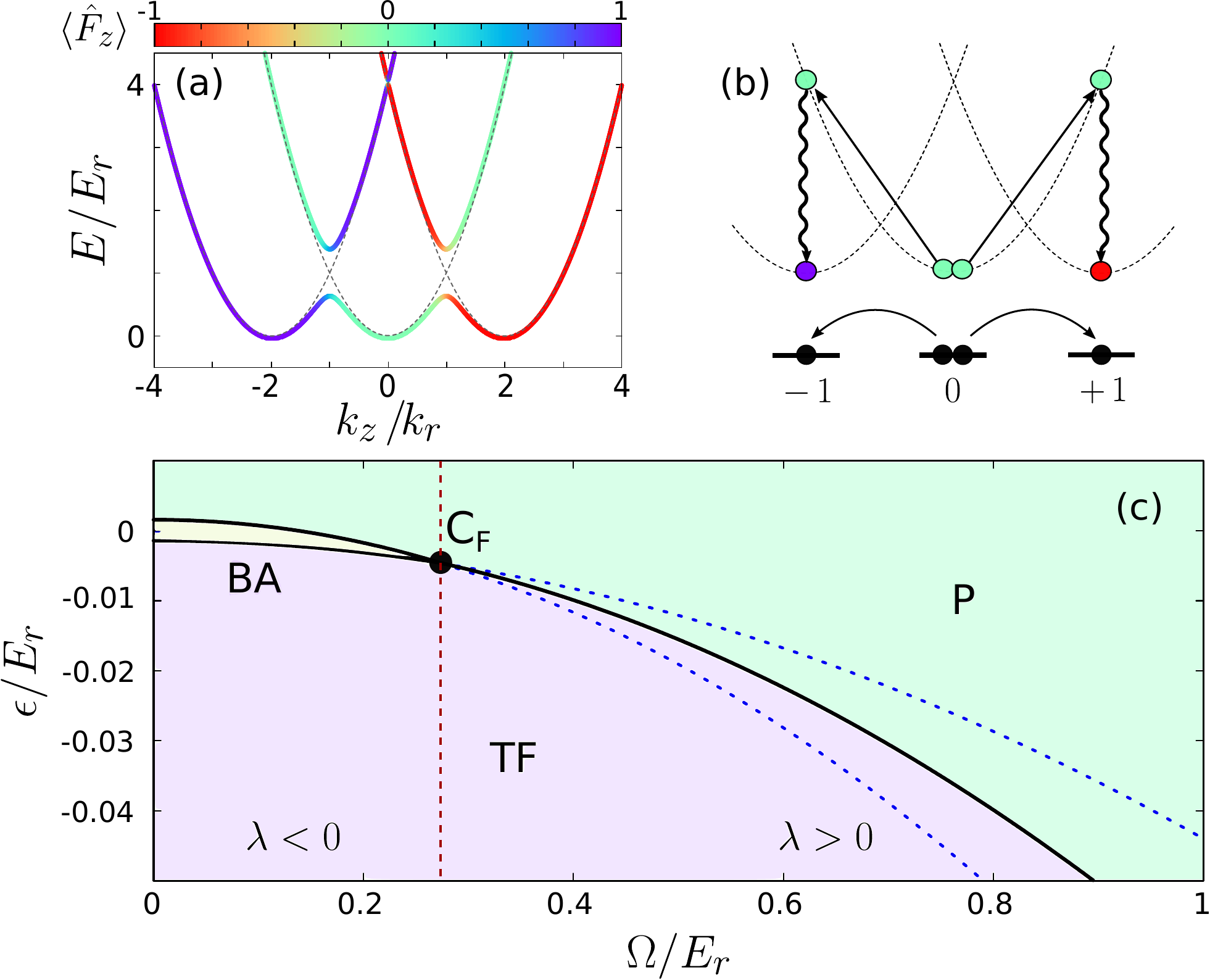}
\caption{
(Color online) \textbf{Pseudospin dynamics in SOC BECs.} (a) Dispersion bands of Hamiltonian \eqref{eq_kinetic_ham}, setting $\Omega= 0.75 E_r$, $\delta=0$ and $\epsilon=\Omega^2/16E_r$. The corresponding mean value of $\hat{F}_z$ for the band states is indicated with the color texture. The undressed bands are shown in dashed gray. (b) Schematic representation of an effective spin-changing collision process enabled by Raman transitions (represented in wavy lines). (c) Phase diagram of the dressed spin Hamiltonian \eqref{eq_spin_ham_mz0}, as a function of the Raman Rabi frequency $\Omega$ and effective quadratic Zeeman shift $\epsilon$, for $^{87}$Rb at $\overbar{n}=7.5\cdot 10^{13}$ cm$^{-3}$. The polar (P), twin-Fock (TF) and broken-axisymmetry (BA) phases meet at the tricritical point $C_F$ (black dot). The dashed red vertical line at $\Omega_c = 4 E_r\sqrt{\protect\abs{g_2}/g_0}$ separates the ferromagnetic ($\lambda <0$) and the antiferromagnetic ($\lambda >0$) regimes of the effective Hamiltonian. The blue dotted lines enclose the region of parameters around the P-TF transition where the BA' excited-state quantum phase takes place (see \sref{Sec-ESphase}), with its boundaries located at $\tilde{\epsilon} = \pm 2\lambda$ in the thermodynamic limit.
}\label{Fig_eff_low_en_phys} 
\end{figure}

In this work, we focus on the weak Raman coupling regime, where $\Omega$ is smaller than the Raman single-photon recoil energy $E_r$.  We label the recoil momentum as $\hbar k_r$, so that $E_r = \frac{\hbar^2 k_r^2}{2m}$. Furthermore, we will consider $E_r \gg \delta, \epsilon$. In this regime, the lowest dispersion band of $\hat{\mathcal{H}}_{\mathrm{k}}$ has three different minima $\vec{k}_j \sim 2 j k_r \vec{e}_z$, with $j \in \left\{-1,0,1\right\}$, as illustrated in \fref{Fig_eff_low_en_phys}(a). As shown in \cite{Cabedo-stripe-arxiv-2021}, in these conditions the dynamics of the dressed gas can be understood in terms of an effective spinor gas with Raman-mediated spin-changing collisions (see \fref{Fig_eff_low_en_phys}(b)). For small condensates, the low-energy landscape of the weakly-coupled gas can be restricted to just three self-consistent modes, and the system is then well described by the following collective pseudo-spin Hamiltonian
\begin{equation}\label{eq_eff_spin_ham}
\ham_\mathrm{eff} = \frac{\lambda}{2N}\hat{L}^2 - \frac{\lambda-{g}_2n}{2N}\hat{L}_z^2 + \delta \hat{L}_z +\tilde{\epsilon}\hat{L}_{zz},
\end{equation}
with the collective pseudo-spin operators $\hat{L}_{x,y,z} = \sum_{\mu\nu}\hat{b}_{\mu}\dgr (\hat{F}_{x,y,z})_{\mu\nu} \hat{b}_{\nu}$ and $\hat{L}_{zz} =\sum_{\mu\nu}\hat{b}_{\mu}\dgr (\hat{F}_{z}^2)_{\mu\nu}\hat{b}_{\nu}$. The bosonic operators 
$\hat{b}_{-1}\dgr$, $\hat{b}_0\dgr$ and $\hat{b}_{1}\dgr$ create a particle in the left, middle and right well mode, respectively. Here, $\lambda = (g_a + g_s\frac{\Omega^2}{16 E_r^2})\overbar{n}$, where $\overbar{n}$ is the mean density of the gas. The coefficient $\tilde{\epsilon}$ includes a perturbative correction to $\epsilon$, with $\tilde{\epsilon} = \epsilon +\frac{\Omega^2}{16 E_r}$. We will restrict ourselves to the the zero ``magnetization'' subspace, where $\hat L_z = 0$. Since $[\ham_\mathrm{eff}, \hat{L}_z ] = 0$, Hamiltonian \eqref{eq_eff_spin_ham} acting on this subspace can be rewritten as
\begin{equation}\label{eq_spin_ham_mz0}
\ham_0 = \lambda\frac{\hat{L}^2}{2N} + \tilde{\epsilon}\hat{L}_{zz}.
\end{equation}

The Hamiltonian \eqref{eq_spin_ham_mz0} is completely equivalent to the one describing the nonlinear coherent spin dynamics of spin-1 BECs where $g_a \ll g_s$ \cite{Law-1998}. Notice that, even in the absence of intrinsic spin-dependent interactions, i.e. $g_a = 0$, Raman dressing enables effective spin-spin interactions, with strength $\lambda \propto \Omega^2$. 
In \fref{Fig_eff_low_en_phys}(c) we plot the phase diagram of Hamiltonian \eqref{eq_spin_ham_mz0} in the $\Omega-\epsilon$ plane using the expression for $\tilde{\epsilon}(\Omega,\epsilon)$ and $\lambda(\Omega)$, and considering a mean density $\overbar{n}=7.5\cdot10^{13}$ cm$^{-3}$ and $g_2/g_0 = -0.0047$. The dashed vertical line at $\Omega = 4 E_r\sqrt{\abs{g_2}/g_0}$ separates the ferromagnetic ($\lambda<0$) and the antiferromagnetic ($\lambda>0$) regimes of the dressed-spin dynamics. The antiferromagnetic regime includes the polar (P) phase at $\tilde{\epsilon}(\Omega) > 0$, in which all the atoms occupy the middle well mode, and the twin-Fock (TF) phase for $\tilde{\epsilon}(\Omega) < 0$, where the atoms evenly occupy both edge-well states. The scenario is richer in the ferromagnetic regime, where the effective spin interactions favor the formation of a non-vanishing transverse magnetization. When the effective interaction dominates, this results in the spontaneous breaking of the SO(2) symmetry of the system \cite{Sadler-2006}, giving rise to the so-called broken-axisymmetry (BA) phase \cite{Murata-2007} in between the P and TF phases. 

\section{ESQPs in SOC gases}\label{Sec-ESphase}

\begin{figure}[t]
\includegraphics[width=0.99\linewidth]{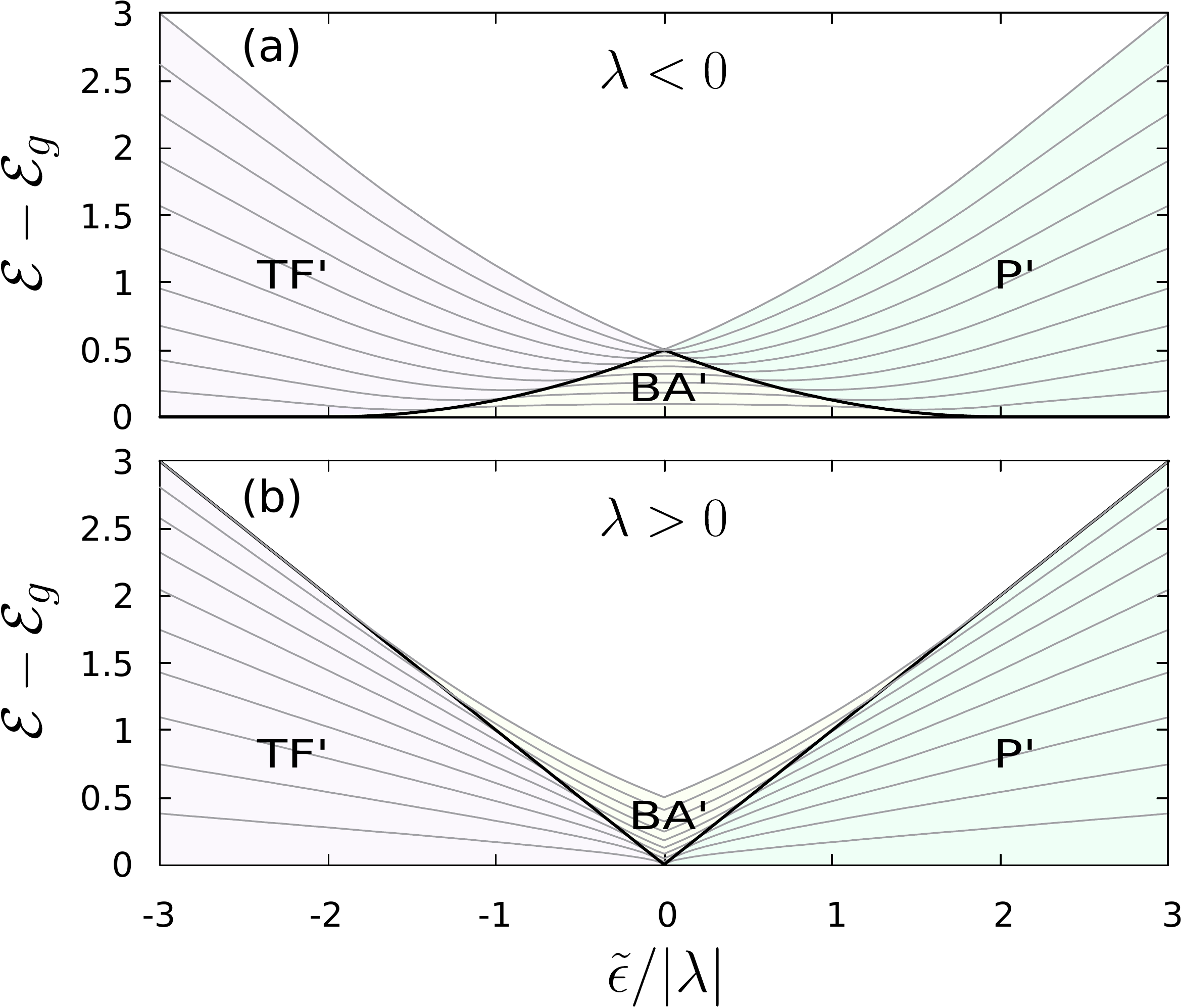}
\caption{(Color online) \textbf{Excited-state quantum phases (ESQPs) in SOC BECs.} ESQP diagram of Hamiltonian \eqref{eq_spin_ham_mz0}, which describe the low-energy landscape of spin-1 gases with SOC coupled, for both effective  ferromagnetic (a) and antiferromagnetic (b) dressed-spin interactions. The thin gray lines show every twenty-fifth eigenvalue of the Hamiltonian for $N=500$. The thick black line indicates the phase boundary
at $\mathcal{E}^* (\tilde{\epsilon})$.}\label{Fig_ESQPs} 
\end{figure}

Ferromagnetic spin-1 BECs, which are described by Hamiltonian \eqref{eq_spin_ham_mz0} with $\lambda<0$, exhibit ESQP transitions \cite{Feldmann-prl-2021}, between three separate ESQPs that extend from the ground state phases and span across the whole energy spectrum. The ESQP diagram of \eqref{eq_spin_ham_mz0} in the $\tilde{\epsilon}-\mathcal{E}$ plane is shown in \fref{Fig_ESQPs}(a) for $\lambda<0$, where $\mathcal{E} = \mean{H}/(\abs{\lambda} N)$ is the scaled energy per particle and ${\cal E}_g$ is the one of the ground state. The phases P', BA' and TF' are labelled according to the corresponding ground state phase. On the boundaries between the phases, the mean-field limit of the density of states diverges, as it is expected for an ESQP transition \cite{Cejnar-jphysA-2021}. The boundaries are found at $\mathcal{E}^* = \tilde{\epsilon}/\abs{\lambda}$ for $-2<\tilde{\epsilon}/\abs{\lambda} <0$, and at $\mathcal{E}^* = 0$ for $0<\tilde{\epsilon}/\abs{\lambda} <2$. Notice that, since $\ham_0 (\lambda, \tilde{\epsilon}) = -\ham_0 (-\lambda, -\tilde{\epsilon})$, the same three phases also occur for antiferromagnetic gases, but with their boundaries redefined, as shown in \fref{Fig_ESQPs}(b), with $\mathcal{E}^* = 0$ for $-2<\tilde{\epsilon}/\abs{\lambda} <0$, and $\mathcal{E}^* = \tilde{\epsilon}/\abs{\lambda}$ for $0<\tilde{\epsilon}/\abs{\lambda} <2$.

As discussed in  \cite{Feldmann-prl-2021}, within these ESQPs the classical phase-space trajectories of coherent states can be classified with respect to a topological order parameter. Here, we show that this order parameter is directly related to the stability of the density modulations in the spin-orbit coupled gas. We exploit this relationship to provide a novel detection protocol for the ESQPs of the spinor gas. 

As in \cite{Feldmann-prl-2021}, we consider now the set of coherent states $\ket{N,\vec{n},\vec{\theta}} = \frac{1}{\sqrt{N!}}\left(\sum_j \sqrt{n_j}\enum{i\theta_j} b_j\dgr \right)^N\ket{0}$ in the zero magnetization subspace, with $\sum_j n_j = 1$ and $n_1 = n_{-1}$. In the mean-field limit of \eqref{eq_spin_ham_mz0}, the scaled energy per particle is given by
%\small
\begin{align}\label{eq_3M_mf_energy_scaled}
\mathcal{E}(\vec{n},\vec{\theta}) &= \bra{N,\vec{n},\vec{\theta}} \ham_0 \ket{N,\vec{n},\vec{\theta}}/\abs{\lambda} N \\%\cr
&= \sgn(\lambda)2(1-n_0)n_0\cos^2\theta + \frac{\tilde{\epsilon}}{\abs{\lambda}}(1-n_0),\nonumber
\end{align}
%\normalsize
where $\theta = \theta_0 - \frac{\theta_1 + \theta_{-1}}2$. The corresponding mean-field equations of motion read
\begin{align}\label{eq_classical_eqs}
\dot{n}_0 &= \frac{\abs{\lambda}}{\hbar}\frac{\partial\mathcal{E}}{\partial \theta}, \qquad \dot{\theta} = -\frac{\abs{\lambda}}{\hbar}\frac{\partial\mathcal{E}}{\partial n_0}.
\end{align}
The solutions of equations \eqref{eq_classical_eqs} are periodic, and the relationship between the periodicity of $n_0(t)$ and $\theta(t)$ varies between the different ESQPs. In the BA' phase, for each point in the $\tilde{\epsilon}-\mathcal{E}$  plane there exist two solutions with disconnected trajectories. In these solutions both $n(t)$ and $\theta(t)$ have the same periodicity. Furthermore, the values that $\theta(t)$ can take are bounded, with $-\pi/2 < \theta(t) < \pi/2$ in one solution and $\pi/2 < \theta(t) < 3\pi/2$ in the other. Conversely, in the P' and TF' phases the solution is unique at each point. Labelling the periodicity in $n(t)$ by $\tau$ , in the P' and TF' phases of the ferromagnetic diagram one has  $\theta(t+\tau) = \theta(t) \pm \pi$. In \cite{Feldmann-prl-2021}, they introduce the winding number
\begin{equation}\label{eq_winding_number}
w = \frac{1}{\pi}\left[\theta(\tau) - \theta(0)\right],
\end{equation}
which can be interpreted as a topological order parameter that distinguishes between the three excited phases. It takes the value $w = -1,0,1$ for any mean-field trajectory within the P', BA' and TF' phases, respectively. In the antiferromagnetic diagram, the sign of $w$ is flipped with respect to the ferromagnetic case. 

\subsection{The Excited-Stripe phase}\label{sub_sec_ES}

Remarkably, we can relate the phase space trajectories $(n_0(t), \theta(t))$ that coherent pseudospin states follow to the properties of the Raman-dressed atomic cloud.  We can write the mean-field wave function of the gas as $\vec{\psi}(\vec{r}) = \sqrt{N} \sum_j \sqrt{n}_j \vec{\phi}_j(\vec{r})\enum{i\theta_{j}}$, where we label the three self-consistent modes around $\vec{k}_j$ as $\vec{\phi}_{j}$. As the three modes are tightly located at the vicinity of the respective band minima $\vec{k}_j$, we can approximate them by plane waves times an slowly varying envelop function, which for simplicity we omit in the following. Then, up to second order in $\Omega/8 E_r$, and neglecting the corrections $\propto (\epsilon+\delta)\Omega^2/E_r^3$, we can write
%\small
\begin{align}\label{planewavefunctions}
\vec{\phi_{1}}(\vec{r}) &\propto \enum{i k_{1}x} \left(1-\frac{1}{2}\left(\frac{\Omega}{8 E_r}\right)^2, \frac{\Omega}{8 E_r}, 0  \right)^{T}, \\%\cr 
\vec{\phi_{0}}(\vec{r}) &\propto \enum{i k_{0}x}\left(\frac{\Omega}{8 E_r}, 1 - \left(\frac{\Omega}{8 E_r}\right)^2 , \frac{\Omega}{8 E_r}  \right)^{T},\cr
\vec{\phi_{-1}}(\vec{r}) &\propto \enum{i k_{-1}x} \left( 0, \frac{\Omega}{8 E_r}, 1-\frac{1}{2}\left(\frac{\Omega}{8 E_r}\right)^2  \right)^{T}.\nonumber
\end{align}
%\normalsize
At $\delta = 0$, $\vec{k}_0 = 0$ and $\vec{k}_{1} = -\vec{k}_{-1}$. In these conditions, the spatial density of the gas reads
\begin{equation}\label{densityofr}
n(\vec{r},t) \sim \overbar{n}\left(1 +  \frac{\Delta n(x,t)}{\overbar{n}}\right),
\end{equation}
where
\begin{align}\label{eq_density_homogeneous_gas}
\Delta n(x,t) &= \overbar{n}\cos(k_1 x - \Delta) \frac{\Omega }{E_r} \sqrt{\frac{n_0(t)(1-n_0(t))}{2}}\cos\theta(t) \cr
&+ O((\Omega/8 E_r)^2).
\end{align}
Here, $\Delta = \theta_{1} - \theta_{-1}$ is the phase difference between the modes at the edge minima, which is a constant of motion at $\delta=0$. In this way, the mean-field solutions of \eqref{eq_spin_ham_mz0} exhibit spatial density modulations that depend both on $n_0$ and $\theta$, with a relative amplitude given by
\begin{equation}\label{eq_amplitude}
A(t) = \frac{\Omega }{E_r} \sqrt{\frac{n_0(t)(1-n_0(t))}{2}}\cos\theta(t). 
\end{equation}
Let us evaluate the behavior of these density modulations in the different phases. In both the P' and TF' excited phases, $n_0(t+\tau)=n(t)$ and $\cos(\theta(t+\tau) = -\cos\theta(t)$. It follows that  
\begin{equation}
\frac{1}{2\tau}\int_{0}^{2\tau} dt  A(t) = 0,
\end{equation}
and so
\begin{equation}
\lim_{T \to \infty}\, \frac{1}{T}\int_{0}^{T} dt  A(t) = 0,
\end{equation}
for all solutions in the P' and TF' phases. Thus, while an excited state in such phases can exhibit spatial density modulations at a given time, such modulations vanish in the time-averaged density profile. 

The situation is different for the BA' phase. There, for each $\tilde{\epsilon}$ and $\mathcal{E}$, one solution fulfills $\cos\theta(t) > 0$ for all $t$ while in the other $\cos\theta(t) < 0$, and thus
\begin{equation}
\lim_{T \to \infty}\abs{\frac{1}{T}\int_{0}^{T} dt  A(t)} > 0.
\end{equation}
Therefore, we can define a new observable that distinguishes a novel ESQP of the SOC spin-1 gas, which we label as Excited-Stripe phase (ES). The classical solutions exhibit a nonzero time-averaged amplitude of the spatial density modulations, or stripes, in the region of parameters that corresponds to the BA' ESQP of the effective dressed spin model of \eqref{eq_spin_ham_mz0}. The topological order parameter $w$ therein is then associated to the stability of the stripes in the Raman dressed spin-1 gas. Such stability is well understood from the locking of the relative spinor phase $\theta$ in the classical mean-field trajectories when $w=0$, which arises from the effective dressed spin-changing collisions in the gas. 

Notice that in presence of a
non-zero detuning $\delta$, the phase of the modulations, $\Delta$, (see eq. \eqref{eq_density_homogeneous_gas}), becomes time dependent, with $\dot{\Delta} = \dot\theta_{1}-\dot\theta_{-1} = 2\delta/\hbar$. While the amplitude of the stripes remains unchanged at leading order, such time dependence of the phase results into vanishing modulations in the time-averaged density profile in the laboratory frame, regardless of the behaviour of $A(t)$. However, there always exist a frame comoving with the modulation where time-averaging of modulations yields the same non-zero value as at $\delta=0$. In practise, the ES phase can be easily distinguished in the presence of non-zero detuning, or even time-dependent, from the behaviour of the contrast of the modulations over time, as discussed in detailed in  \sref{Subsec-signature-ESQP}.

In the ES phase, the contrast of the stripes increases with $\Omega$, and, thus, is larger in the antiferromagnetic regime of \eqref{eq_spin_ham_mz0}, where $\Omega > \Omega_c$. At the same time, for nearly-spin-symmetric gases such as $^{87}$Rb, the region of parameters where the ES can exist is much broader there (indicated with blue-dotted lines in \fref{Fig_eff_low_en_phys}(c)). Yet in this regime the stripe phase does not occur in the ground state of the Raman dressed gas, and one may suspect the gas to undergo a phase separation between the different spin components over time. Still, within the validity of three-mode truncation that leads to \eqref{eq_eff_spin_ham}, phase separation does not occur, and thus the effective model predicts that the stripe phase will persists as excited states even at $\Omega > \Omega_c$ (see \cite{Cabedo-stripe-arxiv-2021}). 

In the next section, we assess by comparison with the mean-field evolution of the whole gas the validity of such truncation, which is equivalent to the single-spatial mode approximation in undressed antiferromagnetic spinor condensates. As the latter, it holds better the smaller the condensate and for zero total magnetization \cite{You-2002-PRA-SMA}. As for the latter, it is notoriously difficult to determine analytically its precise range of validity. Naturally, the physical requirement on the Hamiltonian of the gas for the single-spatial mode approximation to hold is that its non-symmetric part has to be a perturbation of the symmetric part, so that $\lambda \ll g_s \overbar{n}$ and $\lambda \ll \hbar \omega_t$. 

\subsection{The ES phase: Gross-Pitaevskii results}

To verify the predictions of model \eqref{eq_spin_ham_mz0} for Raman dressed SOC gases, we simulate the Gross–Pitaevskii equation (GPE) of the whole system 
\begin{equation}\label{eq_GPE_dressed_gas}
i\hbar \dot{\psi}_j = \delta E[\vec{\psi}]/\delta \psi_j^*,
\end{equation}
where $E[\vec \psi]$ is the energy functional in \eqref{eq_energy_functional}.
We calculate the self-consistent modes $\vec{\phi}_{j}$ via imaginary time evolution of the GPE and define $n_0 = b_0^*b_0$ and $\theta = \arg(b_0)-(\arg(b_{1})+\arg(b_{-1}))/2$, with \begin{equation}
b_j = \frac{1}{N}\int d\vec{r} \vec{\phi}^*_{j}(\vec{r})\cdot\vec{\psi}(\vec{r}).\end{equation}
We consider small $^{87}$Rb condensates in the $F=1$ hyperfine manifold, with $E_r/\hbar = 2\pi \cdot 3678\,$Hz, $k_r = 7.95\cdot 10^{6}$\,m$^{-1}$. We use the corresponding values $a_0 = 101.8 a_B$ and $a_2 = 100.4 a_B$ for the scattering lengths in the different channels, taken from \cite{Stamper-Kurn-2013}, where $a_B$ is the Bohr radius. 

\begin{figure}[t]
\includegraphics[width=0.99\linewidth]{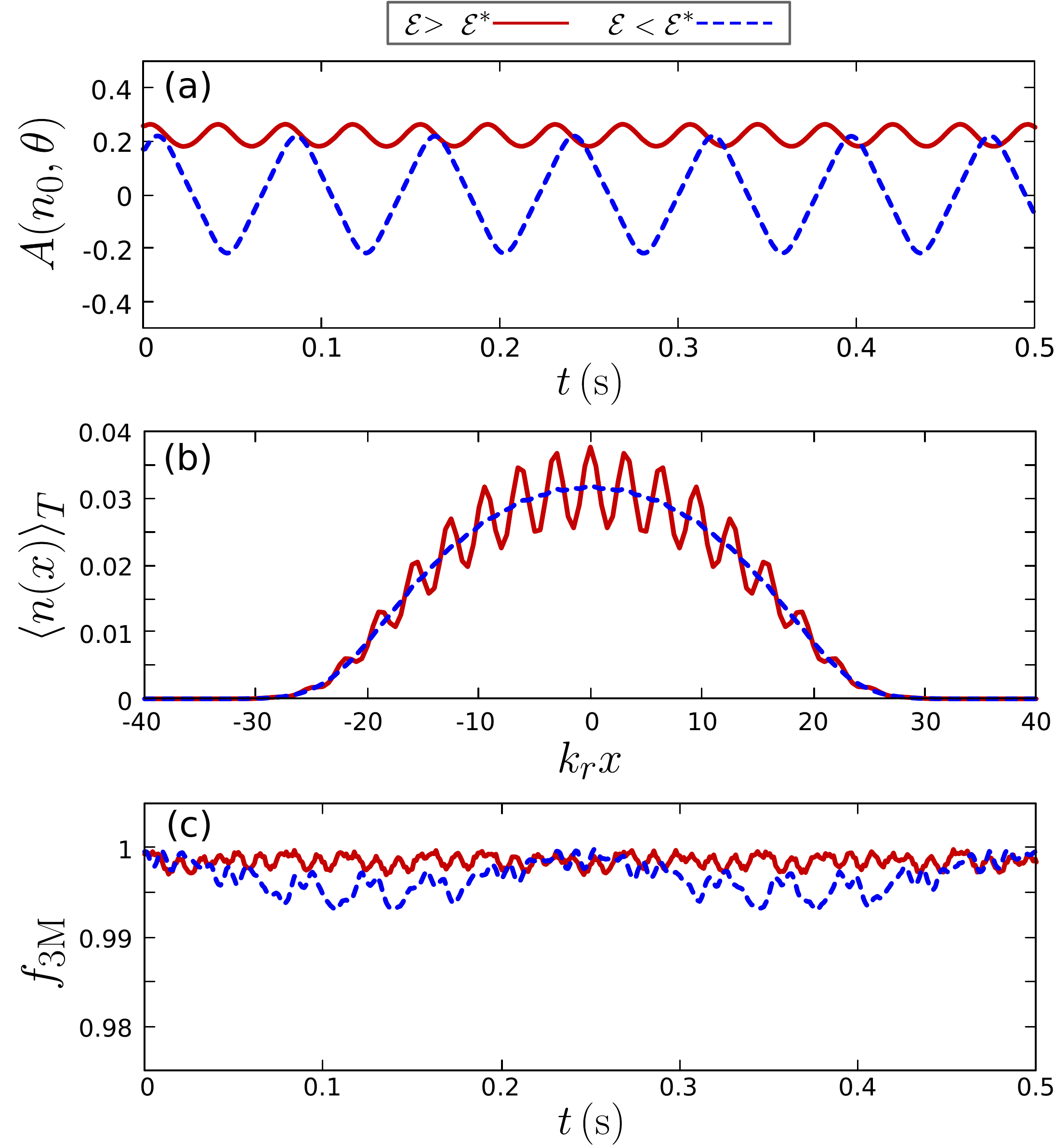}
\caption{(Color online) \textbf{Signature of the excited-state stripe phase.} (a) Relative amplitude $A(t)$ of the spatial modulations for a dressed condensate of $N=10^4$ particles prepared with $\Omega = 0.75 E_r$, $\omega_t = 2\pi\cdot 140$ Hz, $\delta =0$ and $\tilde{\epsilon} = -0.5\protect\abs{\lambda}$, computed using the GPE \eqref{eq_GPE_dressed_gas}. Solid red: $A(t)$ for a state initially at $n_0(0)= 0.5$ and $\theta=0.1 \pi$, with $\mathcal{E} > \mathcal{E}^*$ (ES phase). Dashed blue: $A(t)$ for an initial state at $n_0(0)= 0.5$ and $\theta=0.3\pi$, with $\mathcal{E} < \mathcal{E}^*$ (T' phase). (b) Corresponding time-averaged density profile of the condensate, $\mean{n(x)}_{T}$, averaged over $T=0.5$s. When $\mathcal{E} > \mathcal{E}^*$, the spatial modulation in $\mean{n(x)}_{T}$ does not vanish with increasing $T$. (c) Fraction of the condensate that remains within the subspace spanned by the self-consistent modes $\vec{\phi}_j$.}\label{Fig_init_state_evol} 
\end{figure}

\begin{figure}[t]
\includegraphics[width=0.99\linewidth]{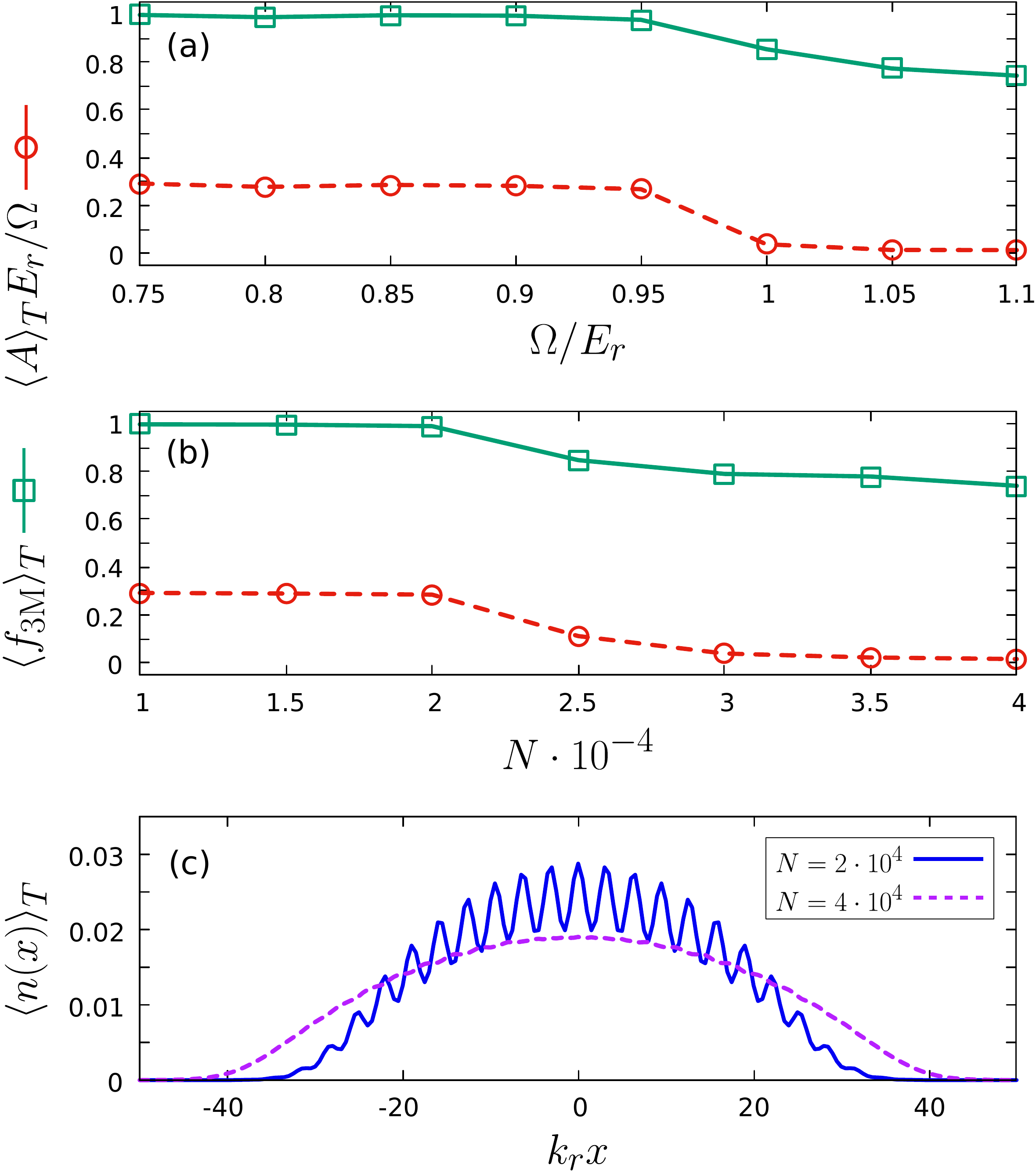}
\caption{(Color online) \textbf{Robustness of the ES phase}
(a) Time-averaged relative amplitude, $\mean{A}_T$, and fidelity of the three-mode truncation, $\mean{f_\mathrm{3M}}_T$, as a function of $\Omega$ for a dressed condensate of $N=10^4$ particles. The state is prepared with $n_0(0)= 0.5$ and $\theta=0.1\pi$, and evolved using \eqref{eq_GPE_dressed_gas} with $\tilde{\epsilon} = -0.5\protect\abs{\lambda}$. (b) $\mean{A}_T$ and $\mean{f_\mathrm{3M}}_T $ as a function of $N$ for a state prepared at $n_0(0)= 0.5$ and $\theta=0.1\pi$, with $\Omega = 0.75 E_r$ and $\tilde{\epsilon} = -0.5\protect\abs{\lambda}$. (c) Time-averaged density profile for the corresponding trajectories with $N=2\cdot 10^4$ and $N=4\cdot 10^4$ from (b). In all cases, the state is evolved for $T = 500\,$ms and $\omega_\mathrm{t}$ is adjusted to have $\overbar{n} = 7.5\cdot10^{13}\,$cm$^{-3}$.}\label{Fig_few_mode_validity} 
\end{figure}

In \fref{Fig_init_state_evol}(a), we plot the relative amplitude $A(t)$ as a function of time for two different states prepared at $\Omega = 0.75 E_r$, $\omega_t = 2\pi\cdot 140$ Hz, and $\delta =0$ with $N=10^4$. In both cases, we adjust $\epsilon$ so that $\tilde{\epsilon} = -0.5\abs{\lambda}$ and set $n_0(0)= 0.5$. We then evolve the initial state with the GPE \eqref{eq_GPE_dressed_gas}. In one trajectory (in solid red), the state is initialized at $\theta=0.1 \pi$, with a corresponding $\mathcal{E} > \mathcal{E}^*=0$, and thus expected to be in the ES phase. Indeed, in agreement with the effective model, $A(t)$ is periodic and remains positive (or negative) at any time $t$, due to the spinor phase being bounded along the mean-field trajectory. Conversely, the dashed blue line corresponds to a trajectory with $\theta(0) = 0.3\pi$, and so $\mathcal{E} < \mathcal{E}^*$, thus out of the ES phase (see \fref{Fig_ESQPs}(b)). In this case the amplitude oscillates between positive and negative values, averaging to $0$ over a period. In \fref{Fig_init_state_evol}(b) we show the corresponding time-averaged density profile of the condensate, given by
\begin{equation}
\mean{n(x)}_{T} = \frac{1}{T}\int_{t_0}^{t_0+T} dt \int dy dz \abs{\vec{\psi}(\vec{r})}^2,
\end{equation}
and averaged over a time $T = 500\,$ms. As expected, $\mean{n}_{T}$ exhibits large modulations when $\mathcal{E} > \mathcal{E}^*=0$, while these vanish for $\mathcal{E} < \mathcal{E}^*=0$. In \fref{Fig_init_state_evol}(c) we plot the fraction of atoms that remain within the three-mode subspace, or fidelity, $f_{3M} = \frac{1}{N^2}\sum_j \left\vert \int d\vec{r} \vec{\phi}_j^*\cdot\vec{\psi}\right\vert^2$, as a function of time, which highlights the accuracy of the approximation in this regime of parameters.

As exemplified by the results shown in \fref{Fig_init_state_evol}, the GPE analysis of the Raman dressed gas supports the predictions of the dressed spin model in a broad, and experimentally accessible, range of parameters. We stress that the stripe phase as an excited-state quantum phase is only well defined and understood within the three-mode subspace, where the robustness of the spatial density modulations is enabled by the collective spin structure of the effective Hamiltonian. The contrast of the modulations in $\mean{n(x)}_{T}$ is very sensitive on the degree of accuracy of the truncation, which in turn depends both on the strength of the effective spin interaction coefficient $\abs{\lambda}$ and on the total number of particles. 

Such sensitivity is illustrated in \fref{Fig_few_mode_validity}, where we show the values of the time-averaged amplitude $\mean{A}_T = \frac{1}{T}\int_{0}^{T} dt A(t)$ and fidelity $\mean{f_\mathrm{3M}}_T = \frac{1}{T}\int_{0}^{T} dt f_\mathrm{3M}(t)$ for a state initialized at $n_0 = 0.5$ and $\theta=0.1 \pi$ and evolved under eq. \eqref{eq_GPE_dressed_gas},  for several values of $\Omega$ in \fref{Fig_few_mode_validity}(a), and for a varying total number of particles in \fref{Fig_few_mode_validity}(b). In all cases, $\omega_\mathrm{t}$ is adjusted so that $\overbar{n} = 7.5\cdot10^{13}\,$cm$^{-3}$, and the quantities are averaged over a total time $T=500\,$ms. In \fref{Fig_few_mode_validity}(a) we set $N=10^4$, and in \fref{Fig_few_mode_validity}(b) $\Omega = 0.75 E_r$. While, according to the effective model $\eqref{eq_spin_ham_mz0}$, the state is prepared within the BA' phase, with $\mathcal{E} > \mathcal{E}^*$, the contrast of the time-averaged density modulations rapidly vanishes as soon as the fidelity of the three-mode truncation degrades. This is exemplified in \fref{Fig_few_mode_validity}(c), where we plot the time-averaged density profile for the corresponding trajectories with $N=2\cdot 10^4$ and $N=4\cdot 10^4$ from \fref{Fig_few_mode_validity}(b). In the latter case, the stripes are absent in the time-averaged density profile, despite having considered the same Raman dressing parameters and atom density than in the former.

It is clear, then, that the collective spin structure is fundamental to the nature of the ES phase. Still, we are able to identify a wide range of parameters for which the few-mode description is accurate, and the behavior of the dressed gas understood in these terms. Furthermore, the direct connection between the ES phase of the Raman dressed gas and the BA' phase of the effective spin model can provide a powerful tool for the detection of the ESQPs of the spinor gas.    

\subsection{Signature of the BA' ESQP}\label{Subsec-signature-ESQP}

In \cite{Feldmann-prl-2021}, the authors propose an experimental scheme to detect the BA' ESQP of a spinor gas. The protocol relies on an interferometric scheme to measure the absolute value of the winding number of \eqref{eq_winding_number}, $\abs{w}$, where the spins are coupled via an internal-state beam splitter after the state is evolved for a period $T$. Such scheme faces a major difficulty: the visibility of the projected measurement is very sensitive to the accumulated phase difference between the $\pm 1$ modes, and hence, to the magnetic field fluctuations in the experiment.

\begin{figure}[t]
\includegraphics[width=0.99\linewidth]{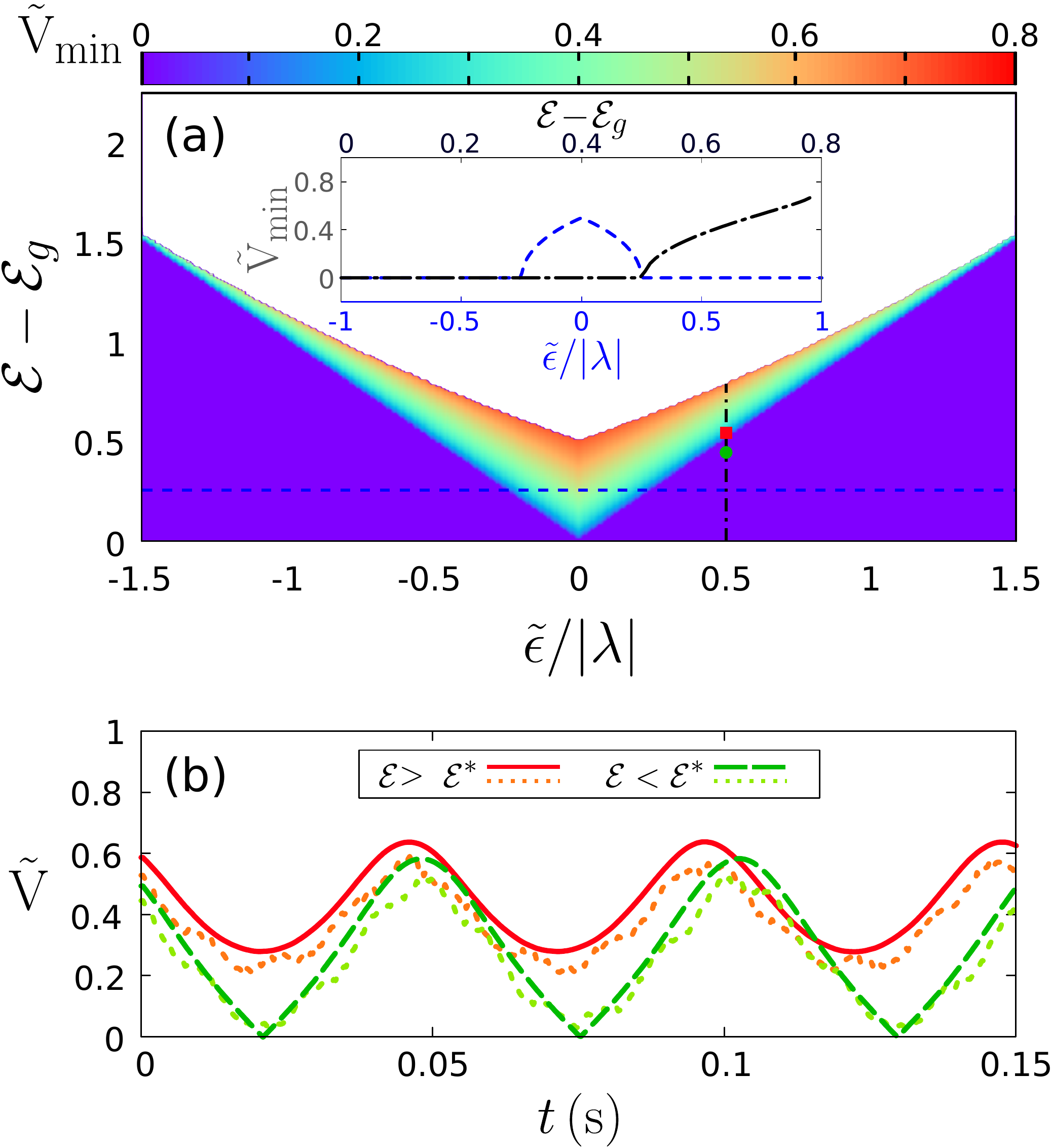}
\caption{(Color online) \textbf{Stripe contrast as signature of the BA' ESQP.} (a) Minimum value $\tilde{\mathrm V}_\text{min}$ of the scaled contrast $\tilde{\mathrm V} = \sqrt{2n_0(1-n_0)}\protect\abs{\cos\theta}$ along the classical trajectories given by equations \eqref{eq_classical_eqs}, as a function of $\tilde{\epsilon}$ and $\mathcal{E}$, computed using eq. \eqref{eq_Vmin}. The inset shows $\tilde {\mathrm V}$ for constant $\mathcal{E}-\mathcal{E}_g = 0.25$ (dashed-blue line) and $\tilde\epsilon /\protect\abs{\lambda} = 0.5$ (dashed-dotted black). (b) $\tilde{\mathrm V}$ as a function of time for two classical trajectories at $\tilde\epsilon/\protect\abs{\lambda} = 0.5$, in and out of the BA'(ES) phase, indicated in (a) by the red and green square dots, respectively. In solid red, $n_0(0) = 0.6$ and $\phi(0) = 0.174$, with $\mathcal{E} > \mathcal{E}^*$. In dashed green, $n_0(0) = 0.6$ and $\phi(0) = 0.243$, with $\mathcal{E} > \mathcal{E}^*$. The corresponding values for the peak-to-valley scaled contrast of the solutions of the GPE \eqref{eq_GPE_dressed_gas} are shown in dotted lines. The values are obtained for a condensate of $N=10^4$ and $\overbar{n}=7.5\cdot 10^{13}\,$cm$^{-3}$, setting $\Omega = 0.75 E_r$. }\label{Fig_ES_visibility} 
\end{figure}

We now show that the realization of the same effective Hamiltonian in the Raman-dressed spinor gas can in principle avoid such drawback. As discussed in \sref{sub_sec_ES}, the amplitude of the spatial density modulations in the dressed gas does not depend at first order in $\Omega/E_r$ on the relative phase $\Delta$, and so neither does the contrast or visibility of the modulations, given by $\mathrm V = 2\abs{A} = (\Omega/E_r)\sqrt{2n_0(1-n_0)}\abs{\cos\theta}$. We conveniently define the scaled contrast $\tilde{\mathrm{V}}$ as
\begin{equation}\label{eq_contrast_V}
\tilde{\mathrm{V}} = V E_r/\Omega = \sqrt{2n_0(1-n_0)}\abs{\cos\theta}.
\end{equation}
The measurement of the contrast of the stripes involves, therefore, a simultaneous measurement of the population $n_0$ and the phase $\theta$. From the behavior of such contrast alone, we can infer the absolute value of the winding number of \eqref{eq_winding_number}, $\abs{w}$, and, thus, detect the BA' phase of the pseudospin gas --the ES phase of the dressed gas-- regardless of the values taken by $\Delta(t)$. 

%{\bf [AC: we cannot repeat twice the same equation, and in any case write it for $\tilde V^2$! The important relation to notice is that $\tilde V^2 = \sgn(\lambda)({\cal E}-\tilde \epsilon/|\lambda|(1-n_0))$, thus, $\dot{(\tilde V^2)}=\epsilon/\lambda \dot n_0$. Anyway, here we can argue that the minimum can be non-zero only when $\theta$ does not reach $\pi/2+m\pi$, i.e. when we are in the BA' or ES phase.]}
The contrast $\tilde V$ is a positive semidefinite quantity and for generic $n_0$ can reach zero only when $\theta$ reaches $(2k+1)\pi/2$, with $k\in \mathbb{Z}$. This obviously occurs in the P' and TF' phases, where $\theta$ is unbounded, but never occurs in the BA' phase where $|\theta|\le \theta_\text{max}<\pi/2$. Thus, the minimum value $\tilde{\mathrm{V}}_\mathrm{min}$ of the scaled contrast \eqref{eq_contrast_V} is a proxy of $|w|$ as it is nonzero only in the BA' phase, as illustrated in \fref{Fig_ES_visibility}(a) where we plot $\tilde{\mathrm{V}}_\mathrm{min}$ along the classical trajectories as a function of $\tilde{\epsilon}$ and $\mathcal{E}$.   
%From \eqref{eq_3M_mf_energy_scaled},\eqref{eq_classical_eqs} and \eqref{eq_amplitude}, it follows that
%\begin{equation}\label{eq_d_amplitude_dt}
%\frac{d (A^2)}{dt} = -\frac{\tilde{\epsilon}}{\hbar}n_0(1-n_0)\cos\theta\sin\theta \left(\frac{\Omega}{E_r}\right)^2.
%\end{equation}
%For $n_0\neq 0,1$, the extreme points of the visibility occur at $\theta = k\pi/2$, with $k\in \mathbb{Z}$. In the P' and TF' ESQPs, the phase $\theta$ is unbounded, and thus the minima $\mathrm{V}=0$ are found at $\theta = (2k+1)\pi/2$. Contrarily, in the BA' phase $\theta$ is bounded to $\abs{\theta} < \pi/2$, and $V$ is maximal or minimal at $\theta=0$ for any trajectory. We can then use \eqref{eq_3M_mf_energy_scaled} to relate $n_0$ to the values of $\mathcal{E}$ and $\tilde\epsilon$, and recover the minimal visibility for any trajectory in the ESQP diagram. 
%This is shown in \fref{Fig_ES_visibility}(a), where the minimum value $\tilde{\mathrm{V}}_\mathrm{min}$ of the scaled contrast \eqref{eq_contrast_V} along the classical trajectories is plotted as a function of $\tilde{\epsilon}$ and $\mathcal{E}$. 
%The minimum value of the contrast captures well the ESQPT, where 
The onset of $\tilde{\mathrm{V}}_\mathrm{min}$ is found at $\mathcal{E}^*$ (see the inset in \fref{Fig_ES_visibility}(a)). In \fref{Fig_ES_visibility}(b) we plot $\tilde{\mathrm{V}}$ as a function of time along two trajectories at $\tilde\epsilon/\abs{\lambda} = 0.5$. We choose the parameters to have one trajectory within the BA' phase, with $\mathcal{E}$ slightly above $\mathcal{E}^*$, and the other in the TF' phase, with $\mathcal{E} < \mathcal{E}^*$. Finally, in dotted lines we plot the corresponding results from the GPE equation of the dressed and trapped gas \eqref{eq_GPE_dressed_gas}. The contrast is computed from the relative peak-to-valley difference at the central peak of the condensate wave function. We note that values of the minimal contrast shown \fref{Fig_ES_visibility}(a) are obtained analytically using \eqref{eq_contrast_V}. By taking the time derivative of expression \eqref{eq_contrast_V} and using \eqref{eq_classical_eqs}, it is clear that, in the BA' phase, $\tilde{\mathrm{V}}$ can only be minimal (or maximal) at $\theta=0$. We then use \eqref{eq_3M_mf_energy_scaled} and \eqref{eq_contrast_V} with $\theta=0$ to retrieve the analytical expression for $\tilde{\mathrm{V}}_\mathrm{min}(\mathcal{E}, \tilde{\epsilon})$, which reads
%\small
\begin{equation}\label{eq_Vmin}
\tilde{\mathrm{V}}_\mathrm{min} = \sqrt{\mathcal{E} \!-\! 
\frac{\frac{\tilde\epsilon}\lambda(\frac{\tilde\epsilon}\lambda\!+\!2 )+\! \frac{|\tilde\epsilon|}\lambda\sqrt{(\frac{\tilde\epsilon}\lambda \!-\!2 )^2\!-\!8(\mathcal{E}-\frac{\tilde\epsilon}\lambda)}}{4}}.
\end{equation}
%\begin{equation}\label{eq_Vmin}
%\tilde{\mathrm{V}}_\mathrm{min} = \sqrt{\mathcal{E} \!-\! \left(\frac{\tilde\epsilon}\lambda\right)
%\frac{\frac{\tilde\epsilon}\lambda\!+\!2\!+\! \sgn(\tilde\epsilon)\sqrt{(\frac{\tilde\epsilon}\lambda \!-\!2 )^2\!-\!8(\mathcal{E}-\frac{\tilde\epsilon}\lambda)}}{4}}.
%\end{equation}
\normalsize Such derivation, however, assumes that the condensates are perfectly located at the three minima of the dispersion band. The presence of trapping leads to a momentum spread of the wavepaquets, decreasing the actual contrast of the stripes in the cloud. This can be observed in \fref{Fig_ES_visibility}(b), where the peak-to-valley contrast evaluated in the condensate wavefunction is slightly lower than the value predicted by eq. \eqref{eq_contrast_V}. Nonetheless, for relatively small trapping frequencies the behavior of the gas in the distinct ESQPs is qualitatively well described by eq. \eqref{eq_contrast_V}. 

In this way, we have shown that the realization of the collective spin Hamiltonian \eqref{eq_spin_ham_mz0} with a Raman-dressed \emph{artificial} spinor gas can provide an alternative approach to the detection of the ESQP transition therein. In the dressed system, we propose to exploit the built-in interferometer that arises from Raman-dressing, where the three quasimomentum-shifted dressed states can spatially interfere due to their non-zero spin overlap. The behavior of the density modulations arising from such interference signals the value of the topological order parameter that characterizes the BA' phase of the effective spin system introduced in \cite{Feldmann-prl-2021}. Our proposal, thus, does not rely on any external interferometric measurement, which results in an intrinsic robustness to magnetic fluctuations. In such a scheme, the precision to delimit the boundary of the BA' phase is subject to the experimental sensitivity associated to the measurements of the density modulations. Remarkably, $\tilde{\mathrm{V}}_\mathrm{min}$ increases abruptly at the boundary, and the modulations of the ES states remain large at any time of the trajectory even for states close to the transition. This can be understood from the fact that in the classical limit $\tilde{\mathrm{V}}_\mathrm{min}$ is the order parameter of a second order phase transition. From \eqref{eq_Vmin} we can see that its susceptibility diverges as
\begin{equation}
\frac{\partial V}{\partial \mathcal{E}} \simeq \frac{\sqrt{C}}{2} (\mathcal{E}-\mathcal{E}*)^{-1/2},
\end{equation}
where $C = 1 + \frac{\abs{\epsilon/\lambda}}{2-\abs{\epsilon/\lambda}}$.

At the same time, the properties of the stripe phase as an excited-state phase can be exploited to facilitate the accessibility of stripe states in experiments with spinor gases. In the next section, we describe a robust protocol to prepare ES states in a spin-1 spinor gas. 

\section{Quench excitation of ES states via coherent spin-mixing}\label{Sec-quenchES}

\begin{figure}[t]
\includegraphics[width=0.99\linewidth]{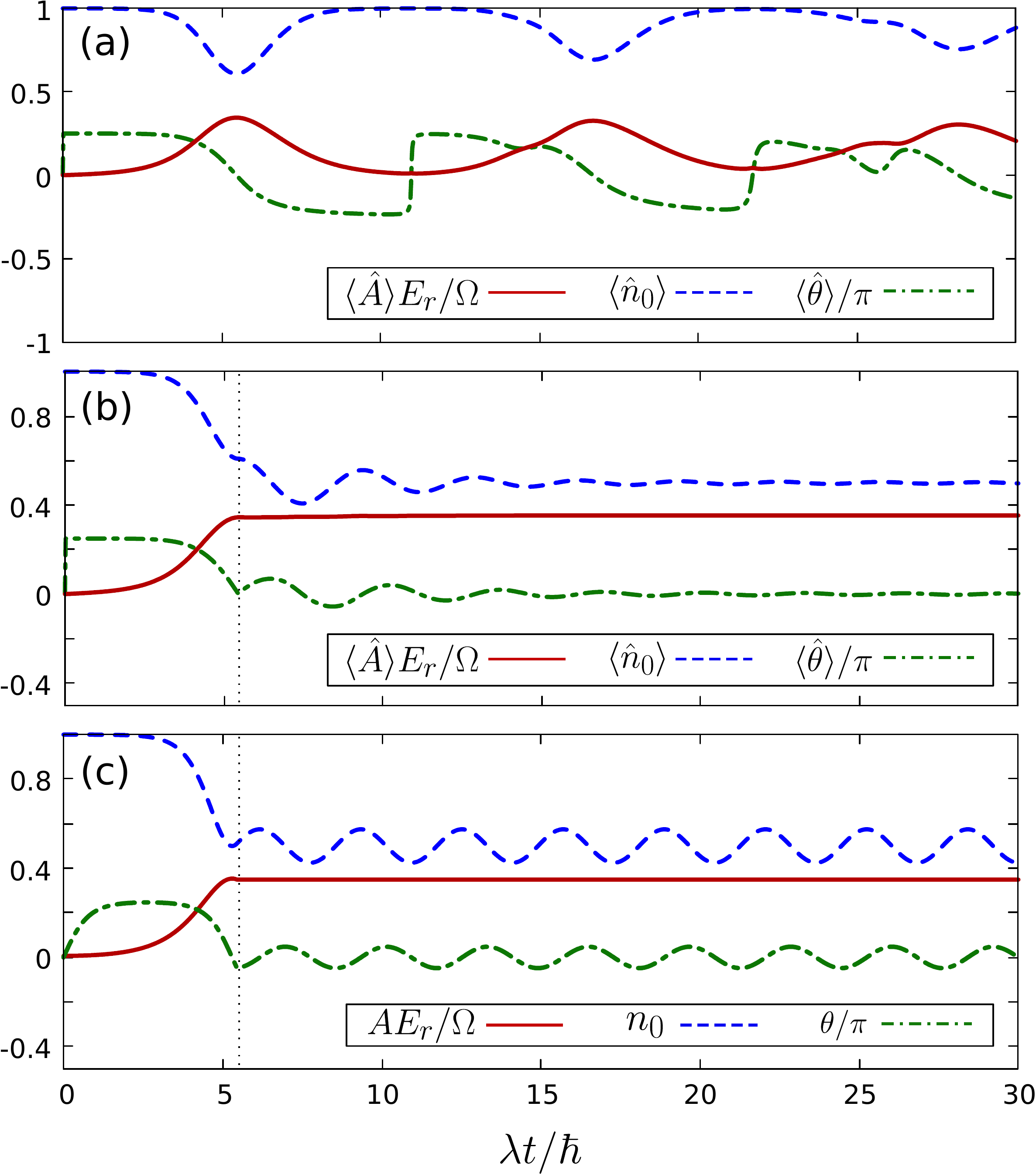}
\caption{(Color online) \textbf{Excitation of ES states via coherent spin mixing: few-mode predictions.} (a) Expected value of $\hat{n}_0$ (dashed blue), $\hat{\theta}$ (dashed-dotted green) and  $\hat{A}$ (solid red) as a function of time for a state prepared at $t=0$ in $\ket{\psi}(t=0) = (\hat{b}_0\dgr)^{N}\ket{0}$, and  evolved under Hamiltonian \eqref{eq_eff_spin_ham}, setting  $\tilde{\epsilon}/\lambda = -1$ and $\lambda > 0$. (b) The same initial state is evolved with $\tilde{\epsilon}/\lambda = -1$ for a time $t_1 = 5.5 \hbar/\lambda$ (indicated with the vertical dotted gray line), where $\tilde\epsilon$ is quenched to $0$. Following the quench, $\mean{\hat{A}}$ stabilizes near its maximum value. (c) Classical trajectories for the state $n_0(0) = 0.9998$  and $\theta(0)=0$ evolved under equations \eqref{eq_classical_eqs}, setting $\tilde{\epsilon}/\lambda = -1$ for $t\leq t_1 = 5.5 \hbar/\lambda$, and $\tilde{\epsilon}= 0$ for $t> t_1$. }\label{Fig_Two_quench_protocol_FQ} 
\end{figure}

Hamiltonian \eqref{eq_spin_ham_mz0} gives a simple framework to understand the collective behavior of SOC condensates. We now use the predictions of the model to propose a protocol that allows a robust and fast preparation of ES states. 
By comparing the rescaled contrast \eqref{eq_contrast_V} and classical energy \eqref{eq_classical_eqs}, we notice that $\tilde V^2 = \sgn(\lambda)({\cal E}-\tilde \epsilon/|\lambda|(1-n_0))$. It immediately follows that at $\tilde{\epsilon} = 0$, $\tilde V$ becomes a constant of motion of the classical trajectories as it is proportional to the square root of the classical energy. With this in mind, we propose a two-step quench scheme to access ES states that exhibit large and stable density modulations. 

\subsection{Two-step quench scheme: few-mode predictions}

We consider that the system is initially in the fully polarized state with $n_0 = 1$, $\theta=0$, where all the atoms occupy the middle-well mode. Experimentally, this scenario is very convenient: we can prepare such state from an undressed polarized condensate in the $m_f=0$ spin state simply by adiabatically turning up $\Omega$ while keeping $\tilde{\epsilon} > 2\abs{\lambda}$. The preparation is followed by a first quench in $\tilde{\epsilon}$ into the range $-2 < \tilde{\epsilon}/\lambda < 0$. According to the classical equations of motion \eqref{eq_classical_eqs}, such polarized state is a stationary point of the Hamiltonian at all values of $\tilde{\epsilon}$. However, quantum fluctuations start a coherent spin-mixing dynamics that breaks the stationarity of the state \cite{Klempt-prl-2010, Dag-2018, evrard-pra-2021-coherent}. In \fref{Fig_Two_quench_protocol_FQ}(a) we show the expected values of the relative occupation of the middle-well mode $\hat{n}_0 = \frac{1}{N}\hat{b}_0\dgr \hat{b}_0$, the spinor phase $\hat{\theta} = \frac{1}{2}\arg\left(\hat{b}_1 \dgr\hat{b}_{-1} \dgr \hat{b}_0\hat{b}_0 \right)$ and the relative amplitude $\hat{A} = (\Omega/E_r)\sqrt{\hat{n}_0(1-\hat{n}_0)/2}\cos(\hat{\theta})$ as a function of time, for the initial state $(\hat{b}_0\dgr)^{N}\ket{0}$ evolved under Hamiltonian \eqref{eq_spin_ham_mz0} with $\tilde{\epsilon}/\lambda = -1$ and $\lambda > 0$. After some time, $\mean{\hat{A}}$ reaches a local maximum. For a coherent state, performing a second quench to $\tilde{\epsilon}=0$ when the maximum is reached would leave $\mean{\hat{A}}$ locked at this value. Naturally, the quantum trajectories of \eqref{eq_spin_ham_mz0} for noncoherent states and away from the thermodynamic limit may depart from the classical predictions. Nonetheless, as expected, we numerically find a qualitative agreement between classical and quantum trajectories, as shown in \fref{Fig_Two_quench_protocol_FQ}(b). In the figure, the initial state $(\hat{b}_0\dgr)^{N}\ket{0}$ is evolved under Hamiltonian \eqref{eq_spin_ham_mz0} with $\tilde{\epsilon}=-\lambda$ for a time $t_1 = 5.5\hbar/\lambda$, where the Hamiltonian is quenched to $\tilde{\epsilon}=0$. Following the second quench, the relative amplitude $\mean{A(t)}$ is rapidly stabilized very near its maximum value $\frac{1}{2\sqrt{2}}\Omega/E_r$. For comparison, in \fref{Fig_Two_quench_protocol_FQ}(c) we show the trajectories obtained using equations \eqref{eq_classical_eqs}. The state is initially in a coherent state with a very small fraction of atoms in the edge well states, to avoid the classical stationary point at $n_0=1$.

\subsection{Excitation of ES states: Gross-Pitaevskii results}

\begin{figure}[t]
\includegraphics[width=0.99\linewidth]{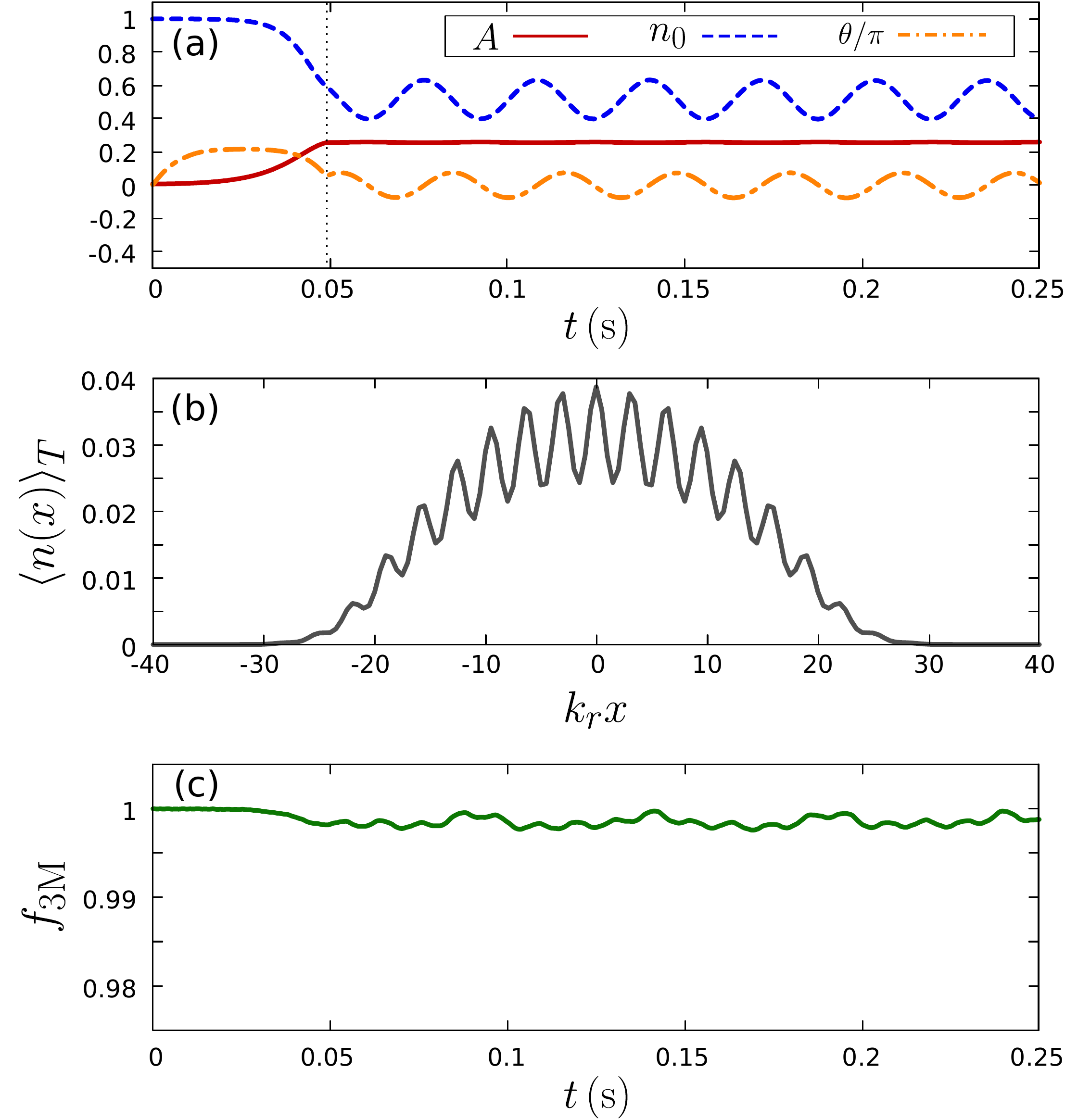}
\caption{(Color online) \textbf{Excitation of ES states via coherent spin mixing: GPE results.} (a) $n_0$ (dashed blue), $\theta$ (dotted-dashed orange) and $A(t)$ (solid red) as a function of time for a state initially prepared at $n_0 = 0.9998$ and $\theta=0$. The state is evolved with the GPE \eqref{eq_GPE_dressed_gas}, for $N=10^4$, $\Omega = 0.75 E_r$ and  $\omega_\mathrm{t}$ adjusted to have $\overbar{n} = 7.5\cdot10^{13}\,$cm$^{-3}$. For $t \leq t_1 = 5.5\hbar/\lambda$, we set $\tilde{\epsilon}/\lambda = -1$. At $t=t_1$ (dotted vertical line) $\tilde{\epsilon}$ is quenched to $0$. (b) Corresponding density profile $\mean{n(x)}_{T}$, time-averaged from $t=t_1$ to $t=0.25\,$s. (c) Relative occupation of the three-mode subspace, $f_\mathrm{3M}$, along the preparation.
}\label{Fig_Two_quench_protocol_GPE} 
\end{figure}

Again, we assess further the validity of the scheme with the GPE of the Raman dressed gas. In order to obtain wide and stable density modulations, we take relatively large values of $\Omega$, and consider small condensates to be safely in the three-mode approximation. \fref{Fig_Two_quench_protocol_GPE} shows a simulation of the protocol with a condensate of $N=10^4$ particles, $\overbar{n} = 7.5\cdot10^{13}\,$cm$^{-3}$ and $\Omega = 0.75 E_r$. In \fref{Fig_Two_quench_protocol_FQ}(a) we plot $n_0$, $\theta$, and $A(t)$ as a function of time for a state initially prepared at $n_0 = 0.9998$ and time-evolved with the GPE. The state is evolved with $\tilde{\epsilon}/\lambda = -1$ for a time $t_1 = 5.5\hbar/\lambda$, where $\tilde{\epsilon}$ is quenched to $0$. As expected, $A(t)$ is stabilized after the quench, despite that $n_0$ and $\theta$ keep oscillating with time. With the contrast stabilized, the time-averaged density profile exhibits very large density modulations, with over 40\% contrast of the stripes, as shown in \fref{Fig_Two_quench_protocol_GPE}(b). In \fref{Fig_Two_quench_protocol_GPE}(c) we plot the values of $f_\mathrm{3M}$ during the evolution, which  remains very close to $1$ for the chosen parameters.

With the two-step quench scheme described, a state with near-maximal density modulations (at a given value of $\Omega$) can be reached in a robust and fast manner. In the example shown in \fref{Fig_Two_quench_protocol_GPE}, $\lambda/\hbar \simeq 2\pi \cdot 17.9\,$Hz, many times larger than the intrinsic spin-mixing rate in a $^{87}$Rb undressed gas. The peak in $A(t)$ is reached in about $50\,$ms. However, the feasibility of the scheme in an actual experiment is subject to the stability of the parameters of the GPE. Several sources of noise can be detrimental to the stability and contrast of the stripes prepared, most notably the fluctuations in the Zeeman levels due to magnetic-field  fluctuations and the calibration uncertainty in the intensity of the Raman beams. We briefly discuss these aspects in the next section.

\begin{figure*}[]
\includegraphics[width=0.9\linewidth]{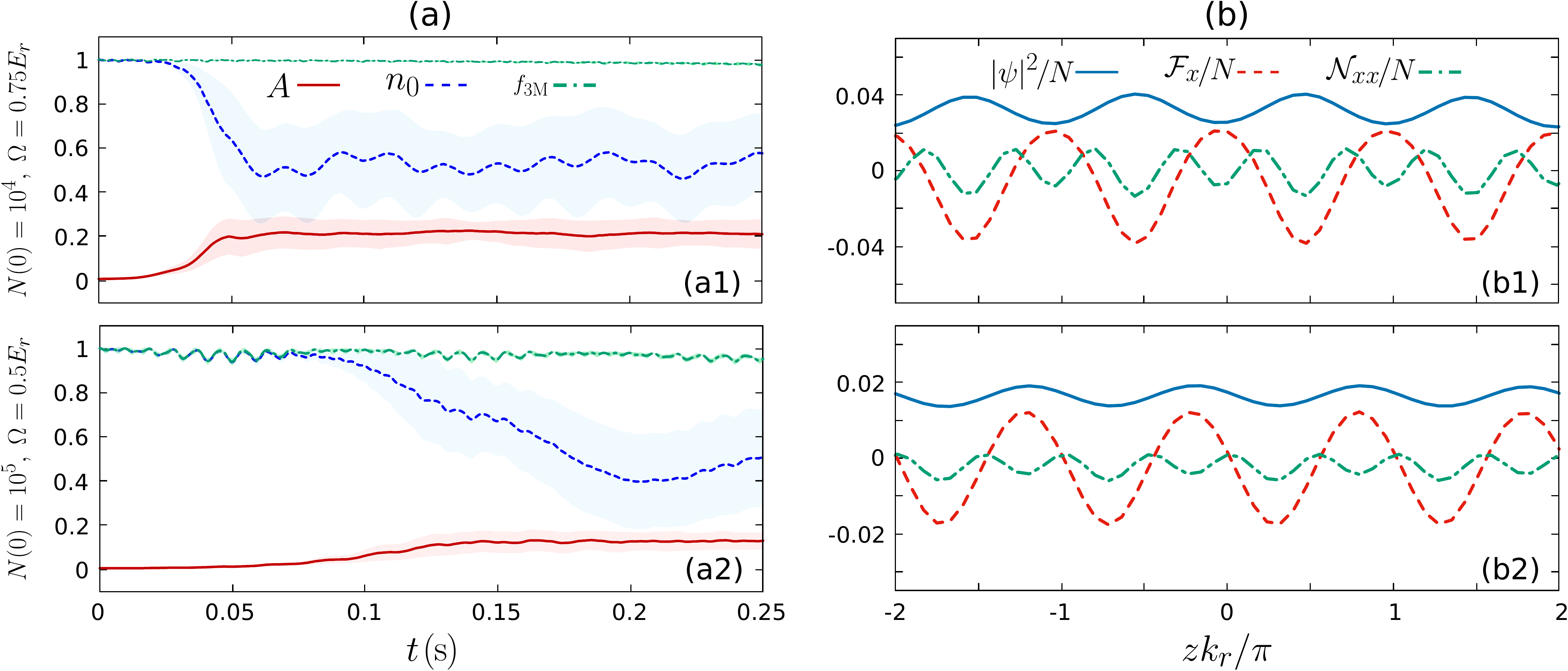}
\caption{(Color online) \textbf{Robust excitation of ES states.} (a) Mean value of $n_0$ (dashed blue) and $A(t)$ (solid red) as a function of time, for a state with $n_0(0) = 0.9998$ and $\theta(0)=0$. The state is evolved under the GPE \eqref{eq_GPE_dressed_gas}, with $N(t)=N(0)\exp(-\gamma t)$ and including randomized parameters to account for atom loss and experimental noise (see main text). In (a1) $N(0)=10^4$ and $\Omega = 0.75 E_r$. In (a2) $N(0)=10^5$ and $\Omega = 0.5 E_r$. In both cases $\gamma=3.33$, and $\tilde{\epsilon}/\lambda = -1$ for $t \leq t_1 = 5.5\hbar/\lambda$ and $\tilde{\epsilon}=0$ for $t>t_1$. The trap frequency $\omega_\mathrm{t}$ is adjusted to have $\overbar{n}(0 = 7.5\cdot10^{13}\,$cm$^{-3}$. The averages are computed from a sample of 40 realizations, with the shadowed regions indicating the associated standard deviation. (b) Longitudinal density $|\vec{\psi}|^2$ (solid blue), spin density $\mathcal{F}_x$ (dashed red) and nematic density $\mathcal{N}_{xx}$ (dashed-dotted green) at $t=t_1$, evaluated for a single realization from the samples used in (a).
}\label{Fig_Two_quench_protocol_GPE_Noise} 
\end{figure*}

\section{Experimental considerations}\label{Sec-experiment}

To benchmark the robustness of the protocol described in \sref{Sec-quenchES}, we include fluctuating and randomized parameters in the simulations of the GPE. To account for atom loss, we continuously renormalize the condensate wave function to $N(t) = N(0)\exp(-\gamma t)$, with $\gamma = 3.33\,$s$^{-1}$, which is compatible with the lifetime of spin-1 Raman-dressed BECs in the considered regimes \cite{Anderson-PRR-2019}. Furthermore, we consider a $10\%$ Gaussian uncertainty in $N(0)$. The background magnetic noise is accounted via sinusoidal modulations of $\delta$ and $\epsilon$ at frequency $50\,$Hz. We set their amplitudes, respectively, to $700$\,Hz and $5$\,Hz, which roughly correspond to a magnetic bias field of $B \sim 35\,$G with $\sim 1$ mG instability in experiments with $F=1$ $^{87}$Rb atoms. We consider a Gaussian uncertainty of $\pm 5\%$ in $\Omega$, to match the systematic uncertainty reported in \cite{Campbell-2016}. Finally, a finite bias field unavoidably results in cross coupling between the two Raman-dressed Zeeman state pairs. This cross coupling is translated into an effective shift in the value of $\epsilon$ that depends on $\Omega$, which can be computed from Floquet theory. We use the polynomial expression for the shift as given in Methods from \cite{Campbell-2016}.

With all these considerations, we reproduce the protocol as described in the previous section, incorporating now the uncertainties in the parameters. In \fref{Fig_Two_quench_protocol_GPE_Noise}(a1) we plot the corresponding mean value and standard deviation of $A(t)$, $n_0$ and $f_\mathrm{3M}$ as a function of time, evaluated from a sample of $40$ realizations. Despite the addition of noise, the preparation still yields large and stable modulations in the density profile for the parameters chosen. As discussed in the previous section, the tunability of the Raman-mediated spin-mixing allows the realization of the protocol in larger condensates. This can be achieved by setting a lower $\Omega$ (see \fref{Fig_few_mode_validity}), but at the expense of a smaller contrast of the stripes, as well as of detrimental effects from noise and atom loss. This is exemplified in \fref{Fig_Two_quench_protocol_GPE_Noise}(a2), where we plot the results for an analogous preparation with $N(0)=10^5$ and $\Omega=0.5 E_r$. The trap frequency is adjusted to initially have $\overbar{n}=7.5\cdot10^{13}\,$cm$^{-3}$. While smaller, the amplitude $A(t)$ is stabilized in less than $200\,$ms, with over half the atoms remaining in the condensate.

In \fref{Fig_Two_quench_protocol_GPE_Noise}(b) we plot the longitudinal density $\abs{\vec{\psi}}^2$, the spin density $\mathcal{F}_x= \vec{\psi}^*\hat{F}_{x}\vec{\psi}$ and the nematic density $\mathcal{N}_{xx} = \vec{\psi}^*(2/3-\hat{F}_{x}^2)\vec{\psi} $ at $t=t_1$, right after the quench to $\tilde\epsilon = 0$. The quantities are computed for a randomly chosen realization from the samples used in \fref{Fig_Two_quench_protocol_GPE_Noise}(a). The values shown are not time-averaged since the instability in $\delta$ induces a back-and-forth displacement of the stripes. However, as discussed in \sref{Sec-ESphase}, the width of the stripes remains stable over time, according to \eqref{eq_amplitude}. In the prepared ES states, the periodicity of the spatial modulations match those of the ground-state ferromagnetic stripe phase \cite{Martone-2016}, with the particle density and the spin densities having periodicity $2\pi/\abs{\vec{k_1}} $, and the nematic densities containing harmonic components both with period $2\pi/\abs{\vec{k_1}} $ and $\pi/\abs{\vec{k_1}}$. Remarkably, this preparation of stripe states via crossing an ESQP transition of the effective model compares favorably, both in its robustness and in the contrast achieved, to the quasiadiabatic preparation through a quantum phase transition proposed in \cite{Cabedo-stripe-arxiv-2021}. 

As discussed in \sref{Sec-ESphase}, due to the instability in the relative phase $\Delta$ between the modes $b_{\pm 1}$, positive and negative values of $A(t)$ can not be distinguished experimentally. However, in the states prepared, the contrast of the stripes $V\sim 2\abs{A}$ remains stable over time and does not vanish at any given time, which is the distinct feature of the ES phase. At the same time, such stability provides a direct measurement of the winding number $w$ that characterises the BA' ESQP of the effective spin Hamiltonian. 

\section{Conclusions}\label{Sec-conclusions}

In this work we have studied the emergence of ESQPs in Raman-dressed SOC spin-1 condensates. Following a dressed-base description, the SOC gas can be interpreted as an undressed spinor gas with effective tunable spin-spin interactions. With this in mind, we have directly connected the corresponding ESQPs of the bare spinor gas to those of the Raman-dressed system. Moreover, due to the coupling between internal (spin) and external (motional) degrees of freedom in the presence of SOC, the phases of the dressed condensate exhibit richer features. Most relevantly, a novel ESQP can be defined in the dressed system, the ES phase, where the atomic cloud exhibits stable density modulations that do not vanish over time. The nature of the phase is understood from the topological order parameter that characterizes the ESQPs of the spinor gas in the regime where the system is described by a collective spin Hamiltonian.

We have numerically assessed the predictions of the effective model with simulations of the GPE of the dressed condensate. We find that, indeed, the collective spin structure Hamiltonian plays a fundamental role to the existence of the ES phase, with its signature quickly vanishing when the few-mode truncation that leads to the effective Hamiltonian is significantly challenged. While such sensitivity supposes a restriction to its experimental realization, we have shown that the large tunability of the system allows a wide regime of parameters for which the phase is supported.

At the same time, we have shown that the realization of the spin Hamiltonian in the dressed condensate can be advantageous when it comes to the detection of the ESQP transitions of the system. So far, the proposal to measure the topological order parameter in undressed quantum gases \cite{Feldmann-prl-2021} relies in an interferometric protocol that is very sensitive to magnetic field fluctuations. In contrast, in the Raman-dressed gas, the same information can be obtained from direct measurements of the density profile of the atomic cloud, with an order parameter, the minimum contrast of the spatial modulations, that is insensitive to fluctuations of the bias field. 

Finally, by exploiting the properties of the ES phase, we have proposed a simple scheme to prepare stripe states with large and stable density modulations. We have numerically tested the robustness of such preparation with the GPE, and found it to be feasible in state-of-the-art experiments with Raman-dressed spinor condensates.

\begin{acknowledgments}
We thank L. Tarruell for insightful discussions on experimental aspects of the Raman coupled BEC. We acknowledge support from the Ministerio de Economía y Competividad MINECO (Contract No. FIS2017-86530-P), from the European Union Regional Development Fund within the ERDF Operational Program of Catalunya (project QUASICAT/QuantumCat), and from Generalitat de Catalunya (Contract No. SGR2017-1646). A.C. acknowledges support from the UAB Talent Research program. 
\end{acknowledgments}

\bibliography{main_arXiv}{}
\bibliographystyle{apsrev4-1}

\end{document}